\documentclass[aps,prc,twocolumn,superscriptaddress,showpacs,floatfix]{revtex4-2}

\usepackage{graphicx}  
\usepackage{dcolumn}  
\usepackage{bm}       
\usepackage{amsmath} 
\usepackage{hyperref} 
\usepackage{amssymb}
\usepackage{xcolor}

\begin{document}
\title{Isotope-Resolved Ba and Xe Yields in Actinide Fission and Correlated Heavy--Light Fragment Systematics}
\author{K.~Pomorski}
\affiliation{National Centre for Nuclear Research, Pasteura 7, 02-093 Warsaw, Poland}

\author{A.~Augustyn}
\affiliation{National Centre for Nuclear Research, Pasteura 7, 02-093 Warsaw, Poland}

\author{T.~Cap}
\affiliation{National Centre for Nuclear Research, Pasteura 7, 02-093 Warsaw, Poland}

\author{Y.~J.~Chen}
\affiliation{China Institute of Atomic Energy, Nuclear Data Center, Beijing 102413, China}

\author{M.~Kowal}
\email{m.kowal@ncbj.gov.pl}
\affiliation{National Centre for Nuclear Research, Pasteura 7, 02-093 Warsaw, Poland}

\author{B.~Nerlo-Pomorska}
\affiliation{Department of Theoretical Physics, Maria Curie-Sk{\l}odowska University, 20-031 Lublin, Poland}

\author{M.~Warda}
\affiliation{Department of Theoretical Physics, Maria Curie-Sk{\l}odowska University, 20-031 Lublin, Poland}

\author{Z.~G.~Xiao}
\affiliation{Department of Physics, Tsinghua University, Beijing 100084, China}

\begin{abstract}
Isotope-resolved post-neutron fission yields in the Ba and Xe chains are calculated 
and benchmarked against evaluated reference data, with emphasis on element-resolved 
isotopic chains $Y(N_f)$ at fixed fragment charge $Z$ and on the consistency of 
heavy--light fragment correlations. Calculations are performed within a 
four-dimensional (4D) Langevin framework employing 
Fourier-over-Spheroid shape parametrization. The benchmark covers spontaneous 
fission of selected Cm and Cf isotopes (including $^{244,246}$Cm and $^{250}$Cf) 
as well as neutron-induced fission at thermal and 14-MeV energies for representative 
actinides in the Th--Pu region (including $^{229}$Th, $^{235}$U, $^{239}$Pu, and 
$^{249}$Cf). The dominant neutron-number maxima are reproduced for a large fraction 
of the isotopic chains considered, indicating that the mean charge partition and the 
average neutron content of the main fission channels are described consistently. 
A systematic residual discrepancy is observed in the isotopic widths: the calculated 
yields often fall off too rapidly on the distribution tails, producing distributions 
that are narrower than the evaluated data, most notably for heavy-fragment chains.
\end{abstract}

\maketitle
\section{Introduction}
\label{sec:intro}

A quantitative description of fission observables must simultaneously capture the collective dynamics from saddle to scission and the ensuing statistical de-excitation of nascent fragments. In particular, an internally consistent treatment of fragment isotopic structure---together with correlated yields in mass, charge, total kinetic energy (TKE), and prompt-neutron multiplicity---remains essential for both fundamental studies and applications involving actinide fission.
The precise delineation of the fragment isotopic landscape—defined by the simultaneous determination of the atomic ($Z$) and neutron numbers ($N$)—represents one of the most stringent benchmarks for any theoretical description of nuclear fission. While mass distributions $Y(A)$ have been extensively documented, high-resolution isotopic yields $Y(Z, N)$ remain notably scarce in the literature due to the significant experimental challenges involved in the unambiguous identification of the atomic number for fast-moving fragments. Only advanced facilities that employ inverse kinematics experiments and high-resolution magnetic spectrometry, such as VAMOS \cite{Ramos2018,Ramos2019PRL239U} or SOFIA \cite{Pellereau2017}, have begun to provide these critical data with sufficient precision.

To bridge the gaps in the available experimental data, the theoretical landscape has evolved to include a diverse array of frameworks, ranging from semi-empirical scission-point models to fully microscopic time-dependent approaches. For instance, the GEF model~\cite{Schmidt2016}, based on a semi-empirical treatment of global fission observables, has proven highly successful in reproducing a wide range of data for nuclear applications. More formally, microscopic descriptions such as the Time-Dependent Generator Coordinate Method (TDGCM) \cite{Regnier2016, Verriere2021} offer a rigorous derivation of the fission process from nucleon-nucleon interactions, although they remain computationally demanding for large-scale systematic studies. Consequently, stochastic transport models based on the Langevin equations \cite{Aritomo2014, Usang2019} have emerged as a powerful compromise, providing a robust dynamical description of the saddle-to-scission evolution while maintaining the computational feasibility required for detailed isotopic yield calculations.
The primary objective of this study is to validate the predictive power of the four-dimensional (4D) Langevin framework against the ENDF/B-VIII.0 evaluated nuclear data library~\cite{NNDC}, focusing on Ba and Xe chains at the isotope-resolved level and on the consistency of heavy--light fragment correlations in post-neutron yields.

\section{Methodology}
\label{sec:method}

In this work, we employ the 4D Langevin framework to model 
spontaneous fission as well as neutron-induced fission of actinides, including 
multi-chance fission where successive pre-fission neutron emissions modify the available excitation energy and fissioning system~\cite{Pomorski2024LM,PomorskiXiao2025ISOLDA}. 
The collective evolution governed by Langevin equations is formulated in a deformation 
space spanned by the elongation $c$, left--right asymmetry $a_{3}$, neck variable $a_{4}$, 
and non-axial deformation $\eta$. The Langevin equations are coupled to a Master-type 
equation for modeling the possible emission of neutrons from the excited fissioning 
system prior to scission~\cite{Pomorski2000NPA}. The cooling of the excited primary fission fragments is treated using a statistical evaporation model~\cite{Delagrange1986}.

The driving potential is obtained from a microscopic--macroscopic energy functional using Fourier-over-Spheroid (FoS) shape parametrization~\cite{Pomorski2023FoS}. The macroscopic part is evaluated using the Lublin--Strasbourg Drop (LSD) model~\cite{PomorskiDudek2003LSD}, whereas the microscopic corrections are computed using a Yukawa-folded single-particle potential \cite{Dobrowolski2016YukawaFolded}. This combination provides a flexible yet physically constrained potential-energy landscape for dynamical fission calculations across a broad range of actinide systems. 

In the following presentation of the formalism, we assume that the fissioning nucleus has mass number $A$ and atomic number $Z$ with a specified excitation energy above the ground state. The Langevin trajectory calculations are performed from the outer saddle point to the scission point, where fragment characteristics are determined. Neutron emission prior to reaching the scission point and fragment cooling are described in the last paragraph of this section.

\subsection{Fourier-over-Spheroid (FoS) shape parametrization}
\label{sec:fos}

The collective dynamics is formulated in a four-dimensional deformation space using the Fourier-over-Spheroid (FoS) parametrization, which provides a flexible representation of strongly deformed, necked, and reflection-asymmetric shapes \cite{PomorskiNerlo2023FoSNonax}. 

\begin{figure}[h]
\centering
\includegraphics[scale=0.5]{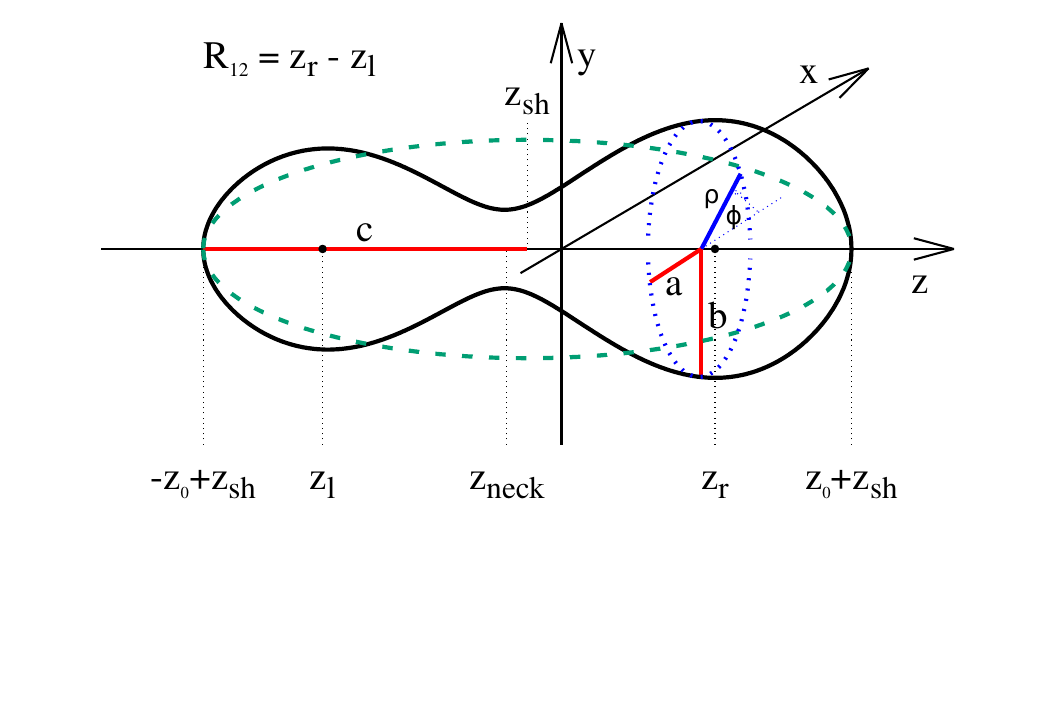}
\caption{Schematic representation of the Fourier-over-Spheroid (FoS) shape parametrization employed in the 4D Langevin description. The nuclear surface (solid black line) is modulated from a reference spheroid (green dashed line). The elongation is controlled by the dimensionless parameter $c$, defining the half-length $z_0 = cR_0$. Non-axiality is introduced via the elliptical cross-section with semi-axes $a$ and $b$, where the surface position is described by cylindrical coordinates $(\rho, \phi, z)$. The axial landmarks $z_l$ and $z_r$ denote the geometric centers of the left and right nascent fragments, respectively, with their relative separation given by $R_{12} = z_r - z_l$. The parameter $z_{\rm neck}$ indicates the location of the minimum neck radius, while $z_{\rm sh}$ represents the axial shift required to maintain the center of mass at the origin. The physical axial boundaries of the shape are defined by $-z_0 + z_{\rm sh}$ and $z_0 + z_{\rm sh}$.}
\label{fig:shape}
\end{figure}

In the present implementation, the elongation is controlled by the dimensionless half-length parameter $c$ (with $z_0=cR_0$, where $R_0$ denotes the spherical-equivalent nuclear radius), while the coefficients $a_3$ and $a_4$ govern reflection asymmetry and necking through a Fourier expansion along the symmetry axis, respectively. Non-axiality is introduced by mapping $(x,y,z)\rightarrow(\rho,\varphi,z)$, allowing for elliptical cross sections perpendicular to the symmetry axis. Figure~\ref{fig:shape} illustrates the FoS geometry used in this work.

\subsection{Non-axial shapes and the $\eta$ deformation}
\label{sec:eta}

To describe non-axial configurations we assume, at a given $z$, an elliptical cross section with semi-axes $a(z)$ and $b(z)$ such that
\begin{equation}
\eta=\frac{b-a}{a+b}, \qquad a(z)b(z)=\rho^2(z),
\label{eq:eta_def}
\end{equation}
where $\rho(z)$ denotes the corresponding axial FoS radius. The parameter $\eta$ plays a role analogous to, but more general than, the classical $\gamma$ deformation in the Bohr parametrization.

The distance from the $z$ axis to the nuclear surface is then written as
\begin{equation}
\rho^2(z,\varphi)=\frac{R_0^2}{c}\,
f\!\left(\frac{z-z_{\rm sh}}{z_0}\right)\,
\frac{1-\eta^2}{1+\eta^2+2\eta\cos(2\varphi)} ,
\label{eq:rho2}
\end{equation}
with the FoS profile function
\begin{equation}
f(u)=1-u^2-\sum_{k=1}^{n}\left[
a_{2k}\cos\!\left(\frac{2k-1}{2}\pi u\right)
+a_{2k+1}\sin(k\pi u)\right].
\label{eq:f_of_u}
\end{equation}
Here $u=(z-z_{\rm sh})/z_0$ with $-1\le u\le 1$, where the shift
\begin{equation}
z_{\rm sh}=-\frac{3}{4\pi}\,z_0\left(a_3-\frac{a_5}{2}+\cdots\right)
\label{eq:zshift}
\end{equation}
ensures that the mass center remains at the coordinate origin. Volume conservation is enforced by the standard FoS constraint,
\begin{equation}
a_2=\frac{a_4}{3}-\frac{a_6}{5}+\cdots .
\label{eq:volume}
\end{equation}
Here, in Eqs.\ref{eq:zshift} and \ref{eq:volume}, higher-order deformations ($a_5$ and $a_6$) not used in this study are also shown for completeness. See details in\cite{PomorskiNerlo2023FoSNonax}.
The above parametrization defines the deformation coordinates used to construct the potential-energy surface and to propagate Langevin trajectories in the collective space.

\subsection{Dissipative collective dynamics: Langevin formalism}
\label{sec:langevin}

The dissipative fission dynamics is formulated in terms of generalized collective
coordinates $\vec q=\{q_i\}=\{c,a_{3},a_{4},\eta\}$ (with $i=1,\dots,4$) and
conjugate momenta $\vec p=\{p_i\}$ through the multidimensional Langevin equations

\begin{equation}
\dot q_i=\sum_{j}\big[{\cal M}^{-1}(\vec q\,)\big]_{ij}\,p_j ,
\label{eq:langevin_q}
\end{equation}
\begin{equation}
\begin{aligned}
\dot p_i={}&-\frac{1}{2}\sum_{j,k}
\frac{\partial\big[{\cal M}^{-1}(\vec q\,)\big]_{jk}}{\partial q_i}\,p_jp_k
-\frac{\partial V(\vec q\,)}{\partial q_i} \\
&-\sum_{j,k}\gamma_{ij}(\vec q\,)\big[{\cal M}^{-1}(\vec q\,)\big]_{jk}\,p_k
+{\cal F}_i(t)\,,
\end{aligned}
\label{eq:langevin_p}
\end{equation}
where ${\cal M}(\vec q\,)$ and $\gamma(\vec q\,)$ denote, respectively, the collective
inertia and friction tensors. In the present implementation, ${\cal M}$ and $\gamma$
are evaluated within the irrotational-flow and wall-type approximations as detailed
in Ref.~\cite{Bartel2019CPC}.

The driving potential is taken as the Helmholtz free energy,
\begin{equation}
V(\vec q\,)=E_{\rm pot}(\vec q,T)-a(\vec q\,)T^2,
\label{eq:free_energy}
\end{equation}
where $T$ is the nuclear temperature and $a(\vec q\,)$ is the deformation-dependent level-density parameter.
This choice provides a consistent thermodynamic measure of the collective drift at finite excitation.

The stochastic force is modeled as
\begin{equation}
{\cal F}_i(t)=\sum_j g_{ij}(\vec q\,)\,\Gamma_j(t),
\label{eq:random_force}
\end{equation}
with $\Gamma_j(t)$ representing independent Gaussian white-noise terms. The diffusion tensor
$D$ is related to friction by the generalized Einstein relation,
\begin{equation}
D_{ij}=\gamma_{ij}\,(T^{*}),
\qquad
D_{ij}=\sum_k g_{ik}g_{kj},
\label{eq:einstein}
\end{equation}
where the effective temperature \begin{equation}
T^{*}=E_0/\tanh(E_0/T) 
\label{eq:tstar}
\end{equation}
accounts for quantum fluctuations \cite{PomorskiHofmann1981}.
The above formulation is used to propagate Langevin trajectories from the outer saddle towards scission in the chosen collective space.

\subsection{Friction strength}
\label{sec:friction}
The collective friction tensor $\gamma_{ij}$ is evaluated within the one-body dissipation framework using the wall formula $\gamma_{ij}^{\rm wall}$ as a basis. A wall mechanism was originally formulated by Błocki and \'{S}wi\c{a}tecki \cite{Swiatecki1978} and further developed by Feldmeier for dissipative heavy-ion dynamics \cite{Feldmeier1987}. This approach describes the transfer of collective energy into intrinsic excitations via nucleon interactions with the moving nuclear surface; however, it is well established that the pure wall formula tends to overestimate the dissipation strength required to accurately reproduce experimental fission observables. To address this overestimation and explore the sensitivity of fragment properties to dissipation, in the present implementation, we employ a phenomenological, temperature-dependent modification based on the formula proposed by Ivanyuk and Pomorski \cite{IvanyukPomorski1996}:
\begin{equation}
\gamma_{ij}^{\rm phen}(T)=
\frac{\tfrac{2}{3}\gamma_{ij}^{\rm wall}}{1+\exp\!\big[(1.3-T)/0.2\big]}+\varepsilon,
\label{eq:gamma_phen}
\end{equation}
where $\varepsilon = 0.01\,\gamma_{\rm wall}$ and $T$ is the nuclear temperature in MeV. This representation provides a practical correction to one-body dissipation, reducing the friction strength to approximately two-thirds of its nominal value at high temperatures. Such a treatment is essential for matching the measured isotopic widths and total kinetic energy (TKE) distributions, as it accounts for the complex interplay between collective motion and intrinsic excitation during the descent from saddle to scission.

\subsection{Temperature dependence of the potential energy}
\label{sec:temp}

Energy-dissipation effects lead to a non-negligible intrinsic excitation already along the descent from saddle to scission, even in spontaneous fission. Temperature effects become particularly important for neutron-induced fission and for compound nuclei formed in heavy-ion reactions. Within the microscopic--macroscopic framework, the deformation- and temperature-dependent potential energy is written as
\begin{equation}
E_{\rm pot}(\vec q,T)=E_{\rm mac}(\vec q,T)+E_{\rm mic}(\vec q,T).
\label{eq:epot_split}
\end{equation}

The macroscopic part is assumed to increase quadratically with temperature,
\begin{equation}
E_{\rm mac}(\vec q,T)=E_{\rm mac}(\vec q,0)+a(\vec q)\,T^2,
\label{eq:emac_T}
\end{equation}
where $E_{\rm mac}(\vec q,0)$ is evaluated using the LSD model \cite{PomorskiDudek2003LSD}. The microscopic contribution $E_{\rm mic}(\vec q,0)$ is computed using a Yukawa-folded single-particle potential~\cite{Dobrowolski2016YukawaFolded}. Shell corrections decrease with increasing temperature, and we employ here the parametrization from~\cite{NerloPomorska2006PRC} where
\begin{equation}
E_{\rm mic}(\vec q,T)\approx
\frac{E_{\rm mic}(\vec q,0)}{1+\exp\!\big[(T-1.5)/0.3\big]},
\label{eq:emic_T}
\end{equation}
with $T$ in MeV. The temperature is determined self-consistently by assuming that the total energy of the system is conserved. Combined with Eq.~\eqref{eq:free_energy}, this yields a deformation-dependent thermodynamic driving potential used in the Langevin propagation.

\subsection{Potential-energy landscape of \texorpdfstring{$^{246}$Cm}{246Cm}}
\label{sec:cm246_pes}

\begin{figure}[h]
\centering
\includegraphics[width=\columnwidth]{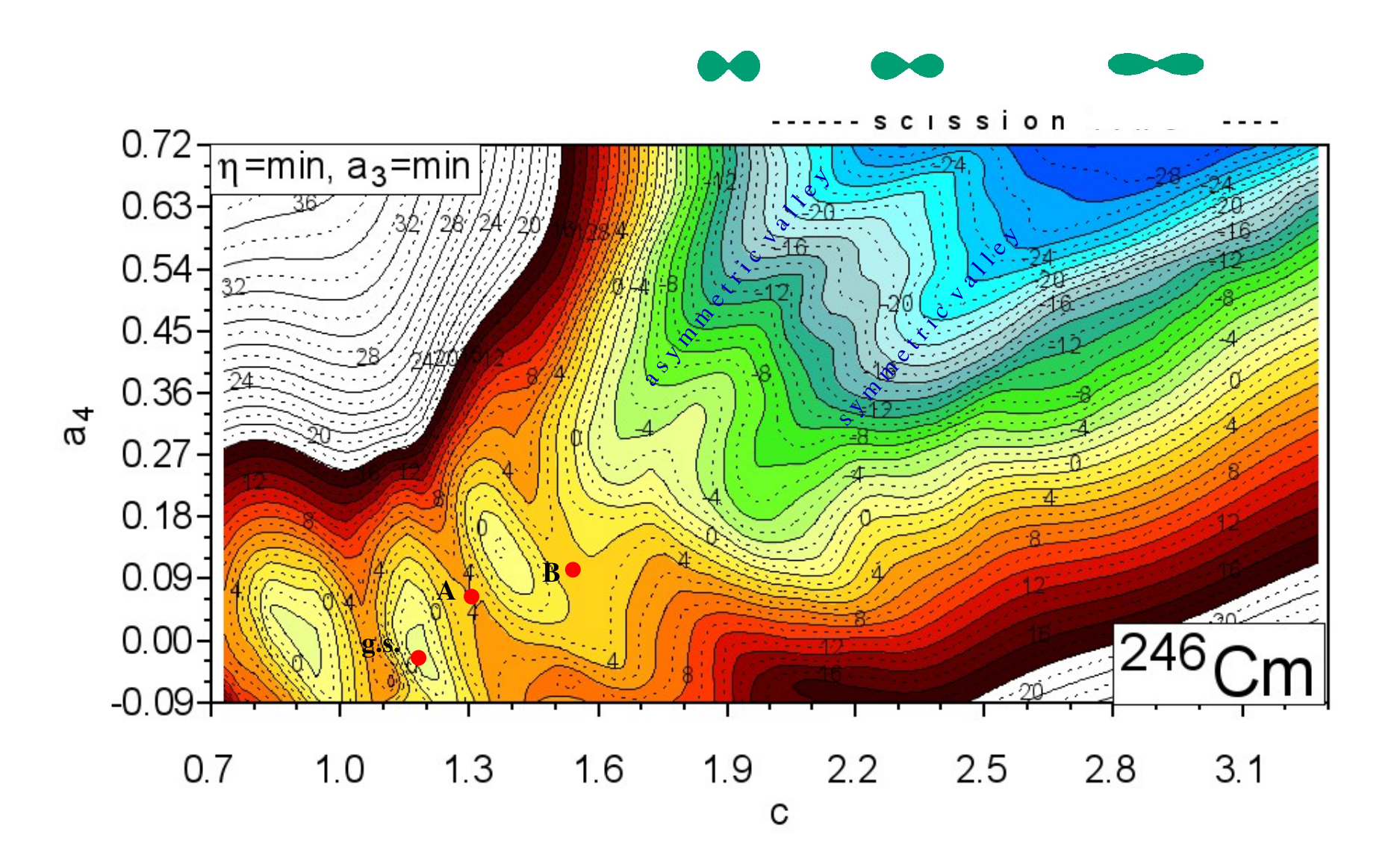}
\caption{Free-energy landscape of $^{246}$Cm in the $(c, a_4)$ plane at $T = 1.4$~MeV, obtained by minimization with respect to $a_3$ and $\eta$. The ``g.s.'' label indicates the ground-state basin. Red points A and B denote the first and second saddle points, respectively. The asymmetric and symmetric descent valleys are separated by a ridge that determines the final mass partition. The dashed ``scission line'' marks the boundary of fragment separation. Green icons at the top symbolize the nuclear shape evolution toward scission.}
\label{fig:cm246_pes}
\end{figure}

The collective dynamics of the fission process are fundamentally governed by the topology of the potential-energy surface. Figure~\ref{fig:cm246_pes} displays the minimized free-energy landscape of $^{246}$Cm in the $(c, a_4)$ plane at a nuclear temperature of $T = 1.4$~MeV. In this representation, the horizontal axis denotes the elongation coordinate $c$, while the vertical axis $a_4$ describes the neck degree of freedom. To obtain this two-dimensional projection, the free energy has been minimized with respect to the remaining degrees of freedom, specifically the reflection asymmetry $a_3$ and non-axiality $\eta$.

The free-energy map provides the key dynamical input for the transport stage: through the drift term $-\partial V(\vec q)/\partial q_i$ the local gradient of the free energy sets the dominant direction of the collective flow, controls the access to competing valleys, and largely determines the scission-point ensemble that seeds fragment formation.

Several critical topological features are explicitly identified on the map. The basin labeled ``g.s.'' corresponds to the ground-state configuration. To progress toward the fission valleys, the system must traverse a double-humped barrier structure characteristic of the actinide region: the point A represents the first saddle point, while point B marks the second saddle point. Beyond the second saddle, the landscape reveals a clear bifurcation into two distinct descent channels—a symmetric valley and an asymmetric valley—separated by a prominent ridge. This structural feature is the primary mechanism governing the partitioning of the probability flux between symmetric and asymmetric mass splits. The upper boundary of the landscape is defined by the dashed ``scission line,'' indicating the approximate locus where a stable neck can no longer be sustained. To assist in interpreting the FoS parametrization, green icons are placed above the map to schematically illustrate the evolution of nuclear shapes—from strongly necked configurations to separated fragments—as the system approaches the scission boundary. 

\subsection{Scission observables: charge partition and total kinetic energy}
\label{sec:scission_obs}

At scission, defined by the endpoint $\vec q_{\rm sc}$ of a Langevin trajectory, the deformation vectors
$\vec q_h$ and $\vec q_l$ are assigned to the heavy and light nascent fragments, respectively. For a fixed
isobaric split $(A_h,A_l)$ with $A_l=A-A_h$, the most probable heavy-fragment charge $Z_h$ is obtained by
minimizing the total scission energy with respect to $Z_h$ \cite{Pomorski2023PRC107},
\begin{equation}
\begin{aligned}
E(Z_h)={}&E_{\rm LSD}(Z-Z_h,\,A-A_h;\vec q_l)
+E_{\rm LSD}(Z_h,\,A_h;\vec q_h) \\
&+V_{\rm Coul}(\vec q_{\rm sc};Z_h,A_h)+V_{\rm nuc}(\vec q_{\rm sc})
-E_{\rm LSD}(Z,\,A;0)\,,
\end{aligned}
\label{eq:charge_energy}
\end{equation}
where $E_{\rm LSD}$ denotes the deformation-dependent macroscopic energy of an $(Z,A)$ nucleus. The Coulomb term at scission is
evaluated as
\begin{equation}
\begin{aligned}
V_{\rm Coul}(\vec q_{\rm sc};Z_h,A_h)=\frac{3e^2}{5r_0}\Bigg[
&\frac{Z^2}{A^{1/3}}\,B_{\rm Coul}(\vec q_{\rm sc})
-\frac{Z_h^2}{A_h^{1/3}}\,B_{\rm Coul}(\vec q_h) \\
&-\frac{Z_l^2}{A_l^{1/3}}\,B_{\rm Coul}(\vec q_l)
\Bigg],
\end{aligned}
\label{eq:coul_scission}
\end{equation}
with $Z_l=Z-Z_h$ and $r_0=1.217$~fm. The short-range interaction at scission is approximated by the surface
energy associated with neck rupture,
\begin{equation}
V_{\rm nuc}(\vec q_{\rm sc})=
-\frac{1}{2}E_{\rm surf}^{\rm sph}\left(\frac{r_{\rm neck}}{R_0}\right)^2, 
\label{eq:vnuc_neck}
\end{equation}
where $E_{\rm surf}^{\rm sph}=b_{\rm surf}A^{2/3}$ is the liquid drop surface energy and $r_{\rm neck}$ is the scission neck radius with $R_0=r_0A^{1/3}$.

Thermal fluctuations of the discrete charge split are incorporated following Ref.~\cite{Pomorski2024LM} by assigning a Wigner-type weight to each integer $Z_i$,
\begin{equation}
W(Z_i)=\exp\left\{-\left[\frac{E(Z_i)-E_{\min}}{T^{*}}\right]^2\right\},
\label{eq:wigner_charge1}
\end{equation}
where $E_{\min}$ is the minimum of $E(Z_i)$ among the sampled integer charges. The effective temperature is taken as in Eq.~\eqref{eq:tstar}. The parameter $E_0$ is taken here around the $\frac{1}{2}\hbar\omega_0$ value, where $\omega_0$ is the nuclear harmonic oscillator frequency. The heavy-fragment charge is sampled from a distribution with partial probabilities defined as
\begin{equation}
P(Z_i)=\frac{W(Z_i)}{\sum_j W(Z_j)},
\label{eq:wigner_charge2}
\end{equation}
and the complementary charge follows from the charge conservation, $Z_l=Z-Z_h$.

The total kinetic energy of the fragments at scission is then evaluated as
\begin{equation}
E^{\rm frag}_{\rm kin}=V_{\rm Coul}(\vec q_{\rm sc})+V_{\rm nuc}(\vec q_{\rm sc})
+E^{\rm coll}_{\rm kin}(\vec q_{\rm sc}),
\label{eq:tke_sum}
\end{equation}
i.e.\ as the sum of Coulomb repulsion, the short-range interaction energy, and the collective pre-scission kinetic energy associated with the relative motion.

Once the mass and atomic numbers of the fragments are determined, their excitation energies are assigned proportionally to the available thermal energy at the scission point, with the proportions calculated according to their level density parameters. Since both fragments are deformed at the moment of scission, we assume that the deformation energy increases the available excitation energy during the cooling stage.

\subsection{Neutron evaporation pre-- and post--scission}
\label{sec:evap}

In the case of spontaneous fission and thermal neutron-induced fission, the available energy is sufficiently small that neutron emission during the system evolution from the ground state to the outer fission saddle can be neglected. In the case of fission induced by neutrons with higher kinetic energies, e.g., 14 MeV, considered in the following section, it is necessary to account for pre-scission neutron emission and consequently multi-chance fission (at moderate excitation energies, charged-particle emission is strongly suppressed). In such cases, the competition between fission and neutron emission is described with a set of coupled Langevin-plus-Master equations, similarly to what has been done in Ref.~\cite{Pomorski2024LM,Pomorski2000NPA}, but using the FoS shape parametrization. For reactions induced by 14 MeV neutrons, second-chance fission dominates. The resulting fragment yield distribution accounts for the proportion of each channel and the changes in excitation energy after each neutron emission.

In the cooling of post-scission fragments, we consider only neutron emission and assume that below the threshold for their emission, further de-excitation proceeds via gamma-ray emission.
The neutron emission width is evaluated within the Weisskopf--Ewing~\cite{Delagrange1986} formalism and is used to determine the energy carried away by each neutron using Monte Carlo methods:
\begin{equation}
\Gamma_{\rm n}(\epsilon)=
\frac{2\mu}{\pi^{2}\hbar^{2}\varrho_{\rm M}(E^{*}_{\rm M})}
\int_{0}^{\epsilon_{\rm n}}
\sigma_{\rm inv}(\epsilon)\,\epsilon\,
\varrho_{\rm D}(E^{*}_{\rm D})\,d\epsilon,
\label{eq:weisskopf}
\end{equation}
where $\mu$ is the neutron reduced mass, $\varrho_{\rm M}$ and $\varrho_{\rm D}$ are the level densities of the mother and daughter nuclei, and $\sigma_{\rm inv}$ is the inverse cross section. The neutron kinetic energy is denoted by $\epsilon$. 
We adopt the Dostrovsky--Fraenkel--Friedlander parametrization \cite{Dostrovsky1959},
\begin{equation}
\sigma_{\rm inv}(\epsilon)=\left[
0.76+\frac{1.93}{A^{1/3}}+
\frac{\tfrac{1.66}{A^{2/3}}-0.050}{\epsilon}
\right]\pi\,(1.70A^{1/3})^{2},
\label{eq:dostrovsky}
\end{equation}
and use the standard Fermi-gas expression for the level density,
\begin{equation}
\varrho(E^{*})=\frac{\sqrt{\pi}}{12\,a^{1/4}(E^{*})^{5/4}}\exp\!\left(2\sqrt{aE^{*}}\right),
\label{eq:fermi_gas}
\end{equation}
with the deformation-dependent level-density parameter $a=a(\vec q\,)$ taken from Ref.~\cite{NerloPomorska2002PRC}. 

This procedure yields prompt-neutron emission characteristics that are consistently coupled to the changes in excitation energy along the fission path.

\section{Results}
\label{sec:results}

We now apply the 4D Langevin framework introduced above to a focused, isotope-resolved test of actinide fission observables and confront the model with element-resolved post-neutron isotopic yields in the Ba and Xe regions. The benchmark covers spontaneous 
fission of selected Cm and Cf isotopes (including $^{244,246}$Cm and $^{250}$Cf) 
as well as neutron-induced fission at thermal and 14-MeV energies for representative 
actinides in the Th--Pu region (including $^{229}$Th, $^{235}$U, $^{239}$Pu, and 
$^{249}$Cf). The analysis proceeds from the heavy-fragment chains to the complementary light partners, thereby testing not only the centroids and widths of individual isotopic distributions but also the heavy-light fragment correlations implied by neutron sharing at scission and later evaporation. This sequence identifies where the model reproduces the data most accurately and where the remaining, typically small discrepancies provide quantitative guidance for further refinement.

\subsection{Discussion: isotope-resolved yields in \texorpdfstring{$^{245}$Cm$(n_{\rm th},f)$}{245Cm(nth,f)}}
\label{sec:disc_cm246}

\begin{figure*}[h]
\centering
\includegraphics[width=\textwidth]{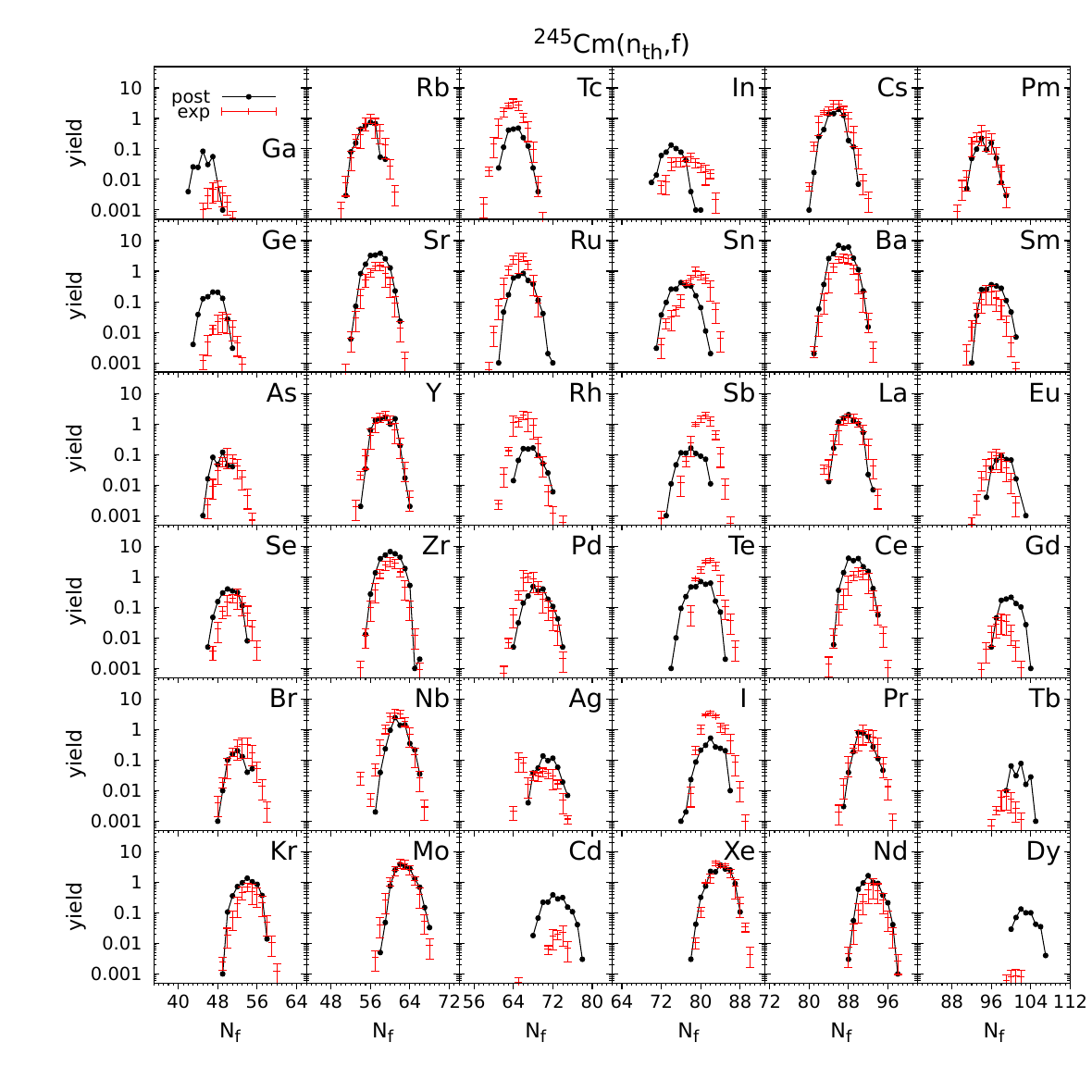}
\caption{Element-resolved post-neutron isotopic yields $Y(N_f)$ for
$^{245}{\rm Cm}(n_{\rm th},f)$. Black symbols and curves represent the model calculations, while red symbols denote the evaluated reference data from ENDF/B-VIII.0~\cite{NNDC}.}
\label{fig:cm246_overview}
\end{figure*}

To provide a clear starting point, we first present in Fig.~\ref{fig:cm246_overview} a compact, system-level comparison in which 36 element-resolved post-neutron isotopic chains are collected in a single panel. This overview for $^{245}$Cm$(n_{\rm th},f)$ allows the two key features of isotope-resolved yields to be assessed at a glance: (i) the position of the maximum in neutron number $N_f$ for each chain (i.e.\ the chain centroid), and (ii) the spread around the maximum, including the effective width and the population of the distribution tails.

As shown in Fig. \ref{fig:cm246_overview}, the agreement is remarkably good over many elements, especially in the vicinity
of the peaks, which is highly encouraging and indicates that the dominant isotopic trends are
captured consistently. The remaining deviations are comparatively small and appear mainly as
subtle differences in the distribution tails of selected chains, providing clear quantitative targets for
further dedicated improvements.

It should also be emphasized that the comparison in Fig.~\ref{fig:cm246_overview} spans an exceptionally
wide range: as indicated by the vertical axis, the yields extend over roughly four to five orders
of magnitude. This alone illustrates the stringency of an isotope-resolved benchmark, because the model is
tested not only at the maxima but also deep into the low-yield regions.

Within this broad range, some element-resolved chains are reproduced with very high accuracy, whereas
others exhibit noticeable deviations. In particular, certain Cd chains show substantial differences from
the reference, while neighboring elements such as Xe and Mo are described much more satisfactorily.
Similarly, Ru and Sb display larger discrepancies, whereas adjacent chains (e.g.\ La) are reproduced
very well. The alternation between well-described and poorly described neighboring elements indicates
that the remaining differences cannot be attributed to a single, obvious deficiency; rather, they likely
reflect a subtle interplay of charge partition, excitation-energy sharing, and neutron evaporation that can
affect specific charge slices differently. These cases, therefore, provide especially informative targets for
further, more focused diagnostics and model refinements.

\begin{figure}[h]
\centering
\includegraphics[width=\columnwidth]{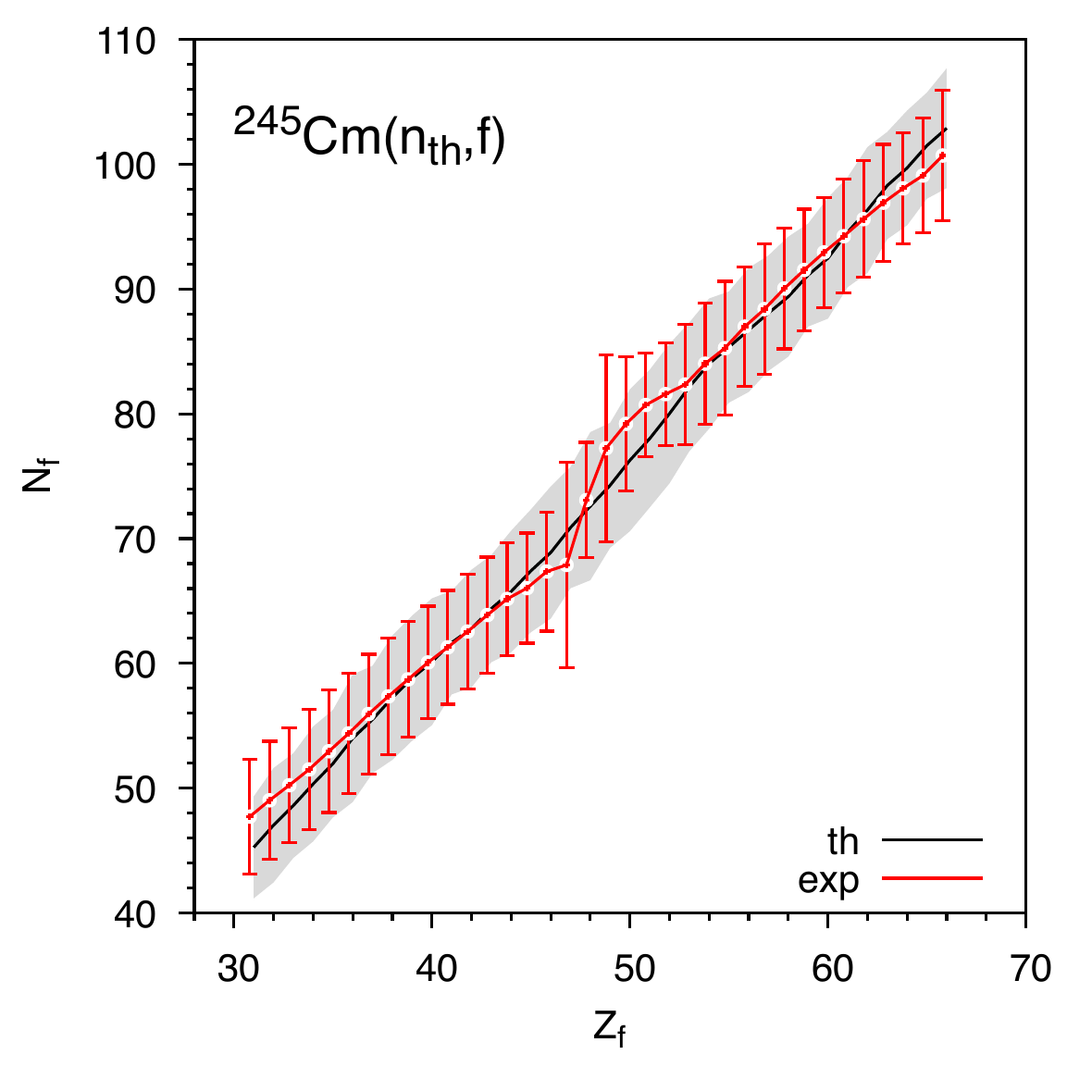}
\caption{Centroid positions of post-neutron isotopic chains as a function of fragment charge 
for the $^{245}{\rm Cm}(n_{\rm th},f)$ reaction. Black line show the calculated 
centroids $\bar{N}_f$ with gray bands representing $3\sigma$ uncertainty ranges. 
Red symbols denote the centroids of evaluated reference data with their corresponding $3\sigma$ 
ranges. All 36 element-resolved chains from Fig.~\ref{fig:cm246_overview} are included.}
\label{fig:centroid}
\end{figure}

To provide a more quantitative assessment of the isotopic-chain systematics, 
Fig.~\ref{fig:centroid} presents the centroid positions as a function of fragment 
charge for all 36 element-resolved chains shown in Fig.~\ref{fig:cm246_overview}. 
The centroid $\bar{N}_f$ and standard deviation $\sigma$ for each isotopic chain 
are computed as
\begin{equation}
\bar{N}_f = \frac{\sum Y_i \cdot N_i}{\sum Y_i}, \quad 
\sigma = \sqrt{\frac{\sum (N_i - \bar{N}_f)^2 \cdot Y_i}{\sum Y_i}},
\end{equation}
where the summation runs over all neutron numbers $N_i$ for a given charge, with 
$Y_i$ being the corresponding yield.

The plot confirms that the model reproduces the overall trend of neutron enrichment 
as a function of fragment charge. However, the largest discrepancies in centroid 
position occur near doubly magic $^{132}$Sn and its correlated light partner. This 
deviation reflects the fact that our current macroscopic charge-equilibration model 
does not explicitly account for shell effects in the fragments---only their deformations 
at scission and odd-even staggering are included. Future calculations will incorporate 
fragment shell corrections more systematically. Additionally, the deviations at both 
ends of the charge distribution suggest that the isospin dependence in the LSD 
macroscopic energy requires refinement, an aspect currently under investigation in the ISOLDA (Iso-scalar liquid drop) framework~\cite{PomorskiXiao2025ISOLDA}.

\subsection{Discussion: Ba and Xe isotopic chains}
\label{sec:disc_ba_xe}

\begin{figure*}[t]
\centering
\includegraphics[width=\textwidth]{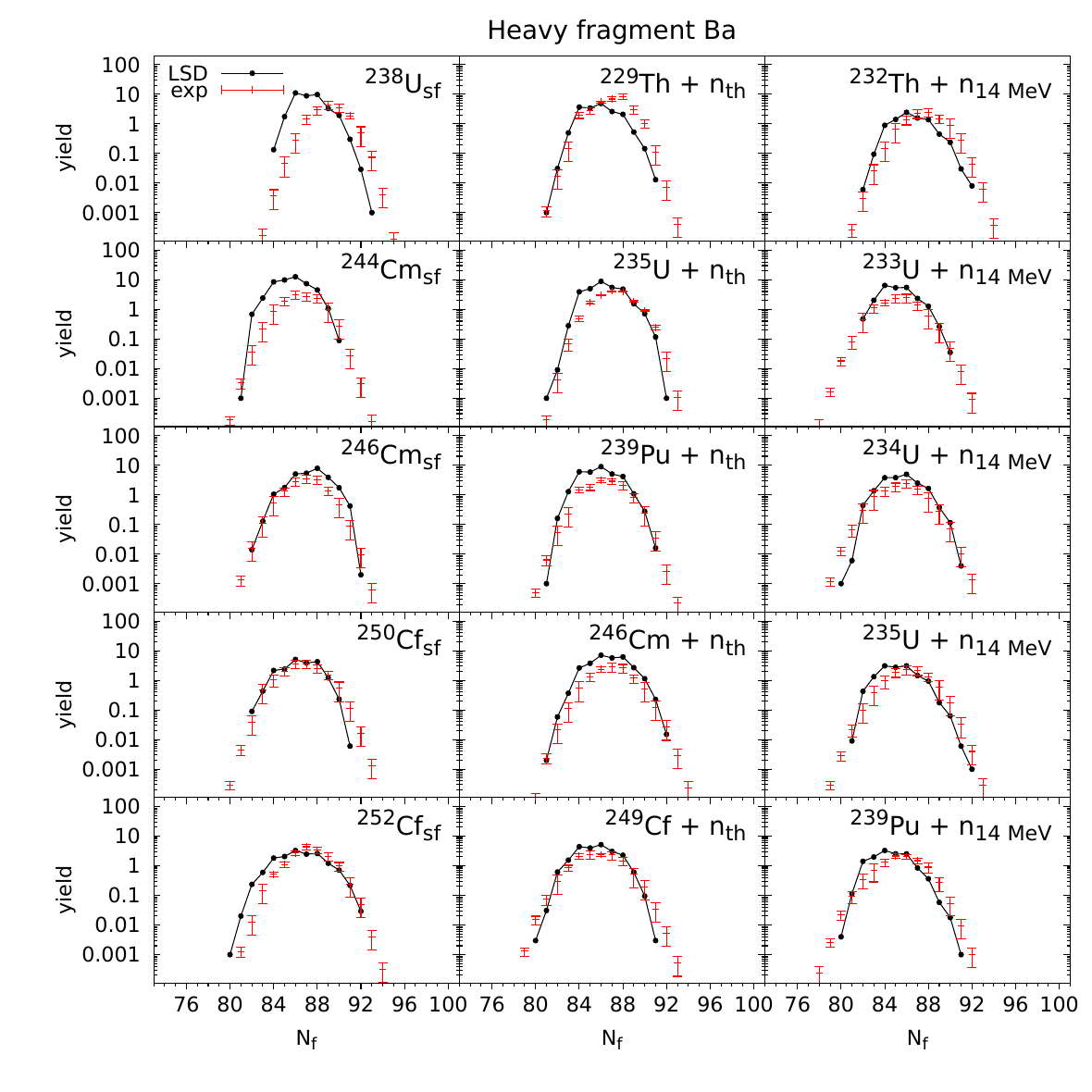}
\caption{Element-resolved post-neutron isotopic yields $Y(N_f)$ for the \emph{heavier} fission fragment with charge $Z=56$ (Ba). Black symbols and curves represent the model calculations, while red symbols denote the evaluated reference data from ENDF/B-VIII.0~\cite{NNDC}.}
\label{fig:heavy_ba}
\end{figure*}

\begin{figure*}[t]
\centering
\includegraphics[width=\textwidth]{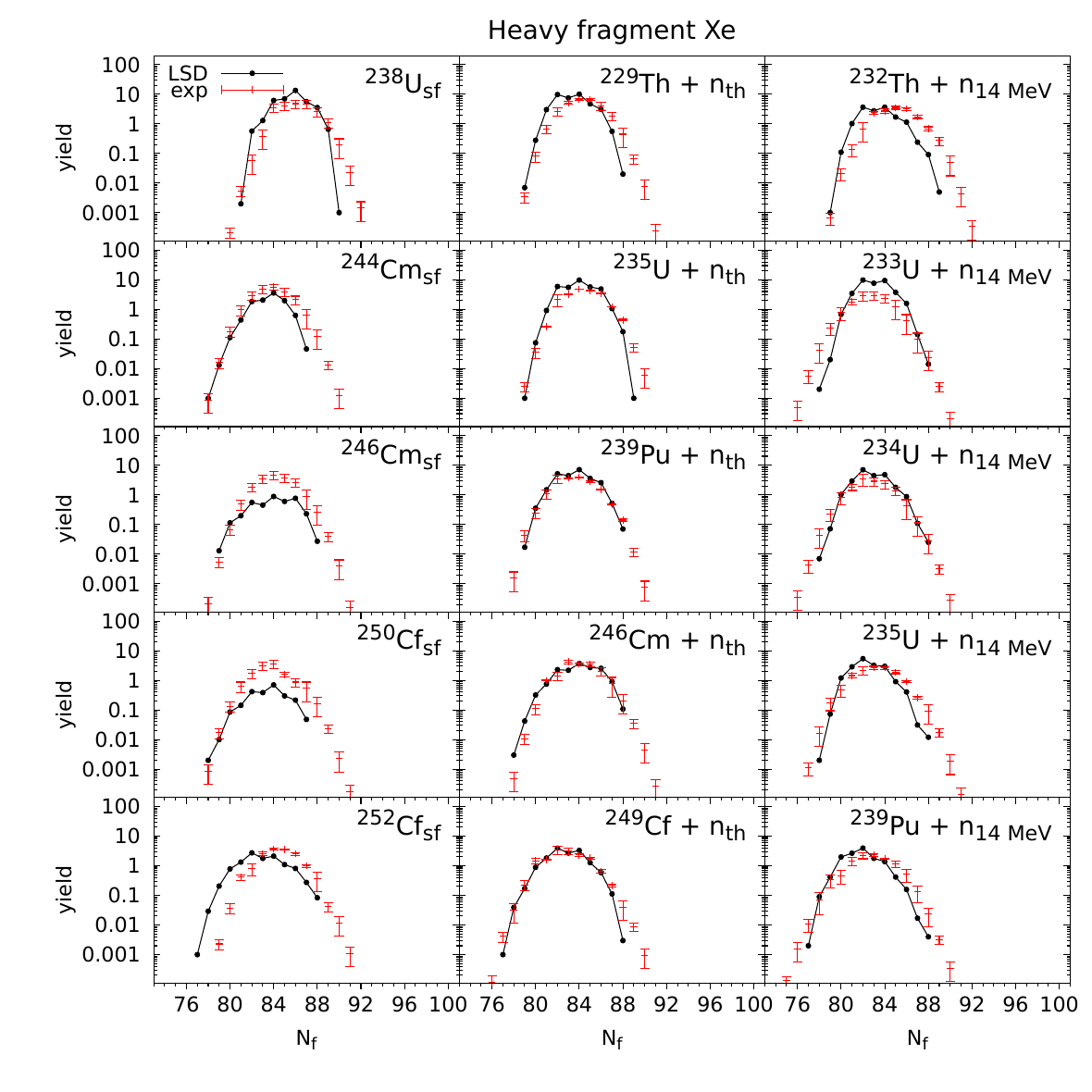}
\caption{Same as Fig.~\ref{fig:heavy_ba}, but the \emph{heavier} fission fragment is Xe ($Z=54$).}
\label{fig:heavy_xe}
\end{figure*}

\begin{figure*}[t]
\centering
\includegraphics[width=\textwidth]{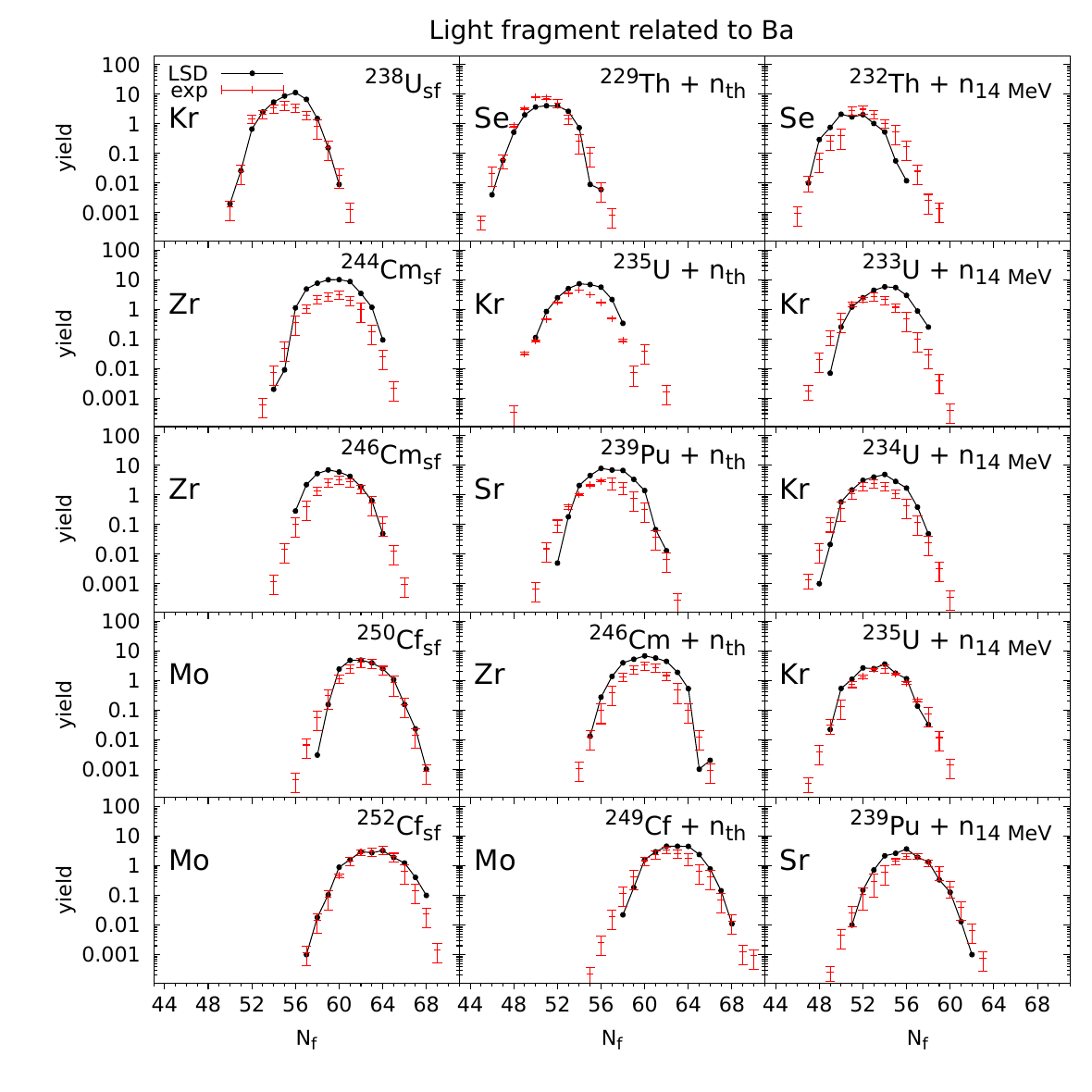}
\caption{Element-resolved post-neutron isotopic yields $Y(N_f)$ for the \emph{light} complementary fragment correlated with the heavy Ba fragment shown in Fig.~\ref{fig:heavy_ba}. Light-fragment element names are indicated in each panel.}
\label{fig:light_ba}
\end{figure*}

\begin{figure*}[t]
\centering
\includegraphics[width=\textwidth]{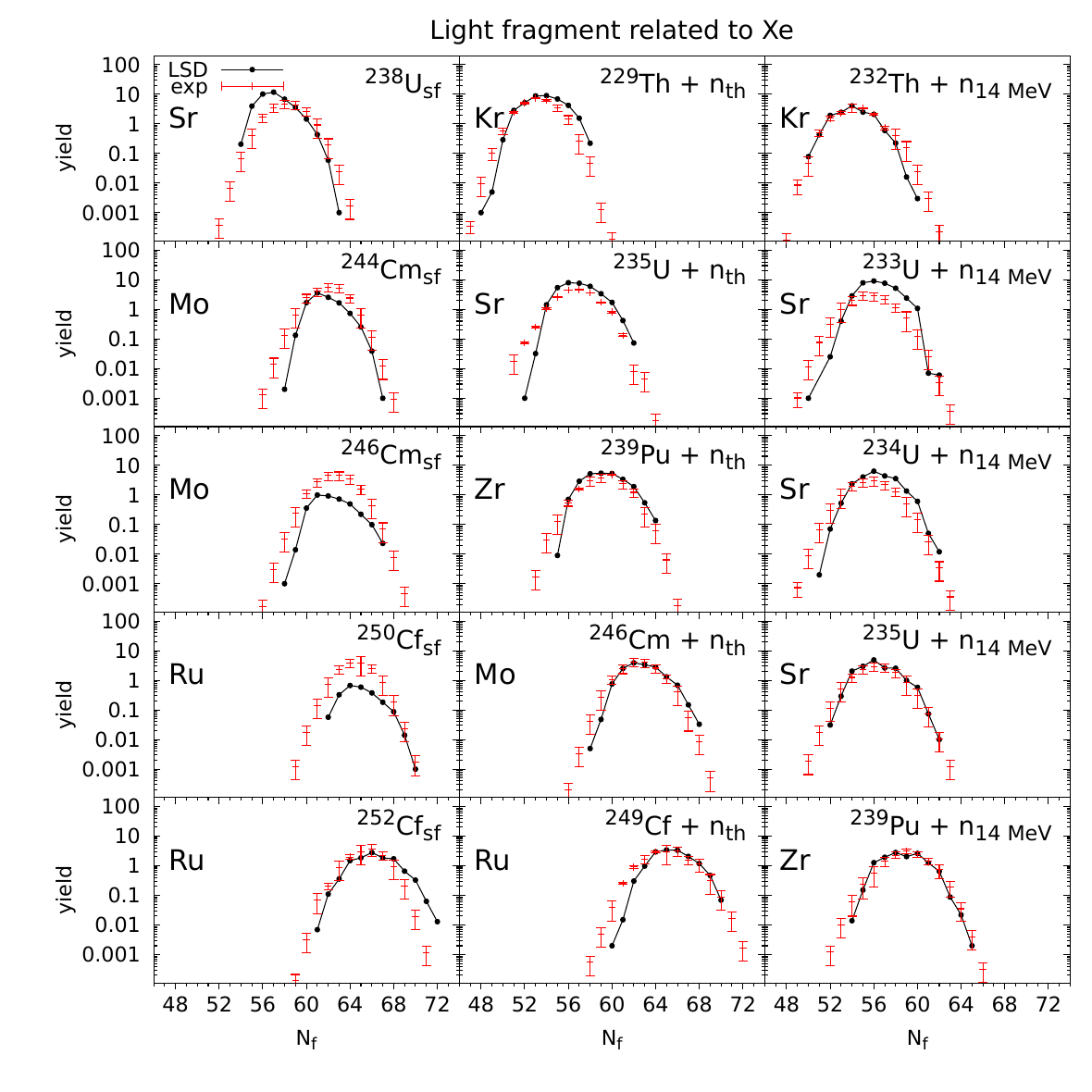}
\caption{Element-resolved post-neutron isotopic yields $Y(N_f)$ for the \emph{light} complementary fragment correlated with the heavy Xe fragment shown in Fig.~\ref{fig:heavy_xe}. Light-fragment element names are indicated in each panel.}
\label{fig:light_xe}
\end{figure*}

Figures~\ref{fig:heavy_ba}--\ref{fig:light_xe} provide a compact but stringent validation of the model at the \emph{isotope-resolved} level for fission of various actinides (both spontaneous and induced). Each panel displays an isotopic chain $Y(N_f)$ at fixed fragment charge $Z$,
with black symbols/curves corresponding to the calculated post-neutron yields and red points representing the
evaluated experimental reference data \cite{NNDC}. In contrast to one-dimensional projections, these element-by-element plots
simultaneously constrain (i) the scission configuration ensemble (mass-asymmetry/charge distribution ),
(ii) the sharing of excitation energy between fragments, and (iii) the subsequent neutron evaporation that
maps primary $(Z,N)$ into the measured $(Z,N_f)$ distributions.

\subsubsection{Heavy-fragment chains: Ba vs Xe}
The isotope systematics for heavy-fragment chains (Ba or Xe) in Figs.~\ref{fig:heavy_ba} and \ref{fig:heavy_xe} show that the model generally reproduces the \emph{centroid systematics} of the heavy peak: for many cases, the position of the dominant maximum in $N_f$ agrees with the evaluated data within the panel resolution, and the near-maximum curvature (the ``core'' of the distribution) is captured credibly. This is particularly visible for a number of neutron-induced cases at low excitation, where the calculated ridge follows the experimental one in a smooth and consistent manner from system to system.

A systematic trend emerges when one inspects the \emph{distribution tails} of the distributions. For a sizable subset of Ba and Xe chains the calculated yields fall off more rapidly than the data, i.e.\ the model produces \emph{narrower isotopic widths} than those inferred from the evaluated data. This effect is most pronounced in panels where the experimental distributions exhibit broad neutron-number dispersion and visible population of long tails. In physical terms, such a narrowing points to an underestimation of the
event-by-event fluctuations that feed a given heavy-fragment isotope chain. Since these widths are shaped by the combined action of (a) diffusion strength along the descent, (b) dispersion of scission excitation energy, and (c) stochastic neutron evaporation, the observed deficit suggests that additional fluctuation strength and/or channel mixing (within the scission ensemble and/or in the de-excitation stage) is required to reproduce the experimental tail yields.

When Ba and Xe are compared directly, the Xe chains (Fig.~\ref{fig:heavy_xe}) are reproduced somewhat more consistently than the Ba chains (Fig.~\ref{fig:heavy_ba}). A likely reason is that the Xe maxima lie closer to the core of the heavy-fragment peak, where the systematics are more strongly constrained by
the dominant fission mode and shell effects; consequently, the centroid position is less sensitive to small
biases in charge partition and excitation-energy sharing. The Ba chains probe a slightly different
charge slice near the same peak and therefore reveal more clearly any residual bias in the predicted neutron
content and, in particular, accentuate differences in the isotopic widths.

\subsubsection{Complementary light fragments}
Figures~\ref{fig:light_ba} and \ref{fig:light_xe} provide an essential \emph{correlation check} for the lighter fission fragment. Because the heavy- and light-fragment isotopic yields are not independent, the simultaneous reproduction of both sides strongly constrains the modeling of neutron sharing at scission and later evaporation. In the best-described cases, the model matches not only the heavy-fragment centroid but also the corresponding light-fragment centroid, indicating that the average partition of neutrons and excitation energy at scission is captured realistically by the model.

Where discrepancies appear, they are informative. In several panels, one observes that the \emph{core} of the light-fragment distribution is reasonably reproduced, but the tails are underpopulated, mirroring the width
deficit seen on the heavy side. This again indicates that the limiting factor is not the gross mass split
itself but the variance generated by fluctuations (dissipative diffusion and/or de-excitation stochasticity).
Conversely, in panels where the light-fragment centroid is slightly displaced while the heavy centroid is
acceptable (or vice versa), the most natural interpretation is a small imbalance in the modeled
charge/neutron distribution  at scission: a mild bias in the predicted $N/Z$ partition will shift the most
probably $N_f$ in opposite directions for the complementary fragments.

A second recurring feature is a weaker odd--even staggering in the calculated yields (where such staggering is visible in the evaluated reference data). In other words, the computed $Y(N_f)$ chains are typically smoother than the data. This is consistent with an implementation that averages over a broad scission ensemble and with a de-excitation treatment that does not fully regenerate odd--even effects to the magnitude indicated by the evaluated data. Cases showing pronounced staggering thus provide a focused benchmark for improving the microscopic ingredients that govern pairing-driven structure during the final stages of fragment formation and neutron evaporation.

\subsection{Model Capabilities and Limitations}
Across the four figures, the model performs best in cases where (i) the experimental isotopic chain is
dominated by a single, well-localized maximum and (ii) the width is moderate, so that the observable is
primarily a centroid/curvature constraint. The most challenging situations are those where the evaluated
data show broad distributions with substantial tail population; there, the calculations tend to be too narrow
for both heavy and light fragments, indicating that the present fluctuation content is insufficient.

Importantly, the fact that a similar width deficit appears in both the Ba- and Xe-gated sets and on both the heavy and light sides suggests a \emph{common origin} rather than an element-specific anomaly. This points to model components shared by all panels: the net diffusion accumulated along the descent, the dispersion of excitation energy at scission, and the stochasticity of neutron evaporation (including its coupling to fragment deformation and level densities). 

Accordingly, Figs.~\ref{fig:heavy_ba}--\ref{fig:light_xe} identify a concrete path for further refinement:
the fluctuation content should be increased in a physically controlled manner while preserving the
well-reproduced centroids, with priority given to the heavy-peak chains, where the outer tails provide the
most stringent diagnostics and where the current deviations are largest.

Generally, one can see that the quality of the reproduction is remarkably good and, in fact, better than one might expect in an isotope-resolved comparison. 
This level of agreement is highly encouraging because it indicates that the dominant mechanisms governing the mean isotopic trends are captured consistently. 
The remaining discrepancies are comparatively minor and appear mainly as subtle differences in the distribution tails, which makes them particularly valuable as quantitative constraints for targeted improvements and further systematic studies.

\section{Conclusions and outlook}
\label{sec:conclusions}

We have presented an isotope-resolved benchmark of post-neutron fission yields in the Ba and Xe regions within the 4D Langevin framework based on Fourier-over-Spheroid shape parametrization. The element-by-element comparison demonstrates that the model reproduces the dominant neutron-number maxima
for a large fraction of the considered isotopic chains, indicating a consistent description of the mean
charge distribution and the average neutron content associated with the main fission channels. This level of
agreement is obtained for spontaneous fission as well as for neutron-induced fission at thermal and 14-MeV
energies, supporting the universal character of the charge-equilibration procedure described in Sec.~\ref{sec:scission_obs}. Importantly, the simultaneous consistency of heavy-fragment and complementary light-fragment systematics provides a non-trivial correlation test of neutron and excitation-energy sharing at scission together with the subsequent neutron evaporation.

The remaining discrepancies are systematic but comparatively small. The most persistent trend is an underestimation of isotopic widths, reflected by a too rapid falloff of yields on the distribution tails, especially for heavy-fragment chains. This points to a slightly insufficient event-by-event dispersion feeding a given $(Z,N_f)$ chain, and identifies fluctuation mechanisms (diffusion accumulated along the descent, scission sampling, and stochastic de-excitation) as the most promising direction for targeted refinement. The Ba and Xe chains emerge as the most stringent constraints because they accentuate the sensitivity to the combined action of charge distribution, excitation-energy sharing, and evaporation.

Overall, the reproduction is remarkably good and highly encouraging; the small residual differences are precisely of the kind that can guide further dedicated studies aimed at improving the width and tail systematics without degrading the well-reproduced centroids and peak structure. Natural next steps include systematic sensitivity studies of the fluctuation content and de-excitation inputs, and an extension of the collective space (e.g.\ higher-order FoS deformations) to quantify possible additional channel mixing effects on isotope-resolved yields.

\begin{acknowledgments}
This research was funded in part by the National Science Centre, Poland, under Project No. 2023/49/B/ST2/01294. It was also supported by the Natural Science Foundation of China under Grant Nos. 11961131010 and 12335008. In addition, M. Kowal and T. Cap were partially supported by the Polish-French cooperation COPIGAL.
\end{acknowledgments}
\bibliographystyle{apsrev4-2}
\bibliography{Bib-KPomorski}

\end{document}